\documentclass[10pt]{article}
\pdfoutput=1

\usepackage{amsmath}
\usepackage{amssymb}

\usepackage{graphicx}

\usepackage{cite}

\usepackage{color}

\usepackage[labelfont=bf,labelsep=period,justification=raggedright]{caption}

\makeatletter
\renewcommand{\@biblabel}[1]{\quad#1.}
\makeatother

\date{}

\usepackage{pdfpages}
\usepackage{hyperref}
\newcommand{\micron}{$\mu\mbox{m}$ }

\begin{document}
\begin{flushleft}
{\Large
\textbf{Virtual {\em in situs}: Sequencing mRNA from cryo-sliced {\em Drosophila} embryos to determine genome-wide spatial patterns of gene expression}
}
\\
Peter A. Combs$^{1,\ast}$,
Michael B. Eisen$^{2,3}$
\\
\bf{1} Graduate Program in Biophysics, University of California, Berkeley, California, United States of America
\\
\bf{2} Department of Molecular and Cell Biology, University of California, Berkeley, California, United States of America
\\
\bf{3} Howard Hughes Medical Institute, University of California, Berkeley, California, United States of America
\\
$\ast$ E-mail: peter.combs@berkeley.edu
\end{flushleft}

\section*{Abstract}

Complex spatial and temporal patterns of gene expression underlie embryo differentiation, yet methods do not yet exist for the efficient genome-wide determination of spatial expression patterns during development. {\em In situ} imaging of transcripts and proteins is the gold-standard, but it is difficult and time consuming to apply to an entire genome, even when highly automated. Sequencing, in contrast, is fast and genome-wide, but is generally applied to homogenized tissues, thereby discarding spatial information. It is likely that these methods will ultimately converge, and we will be able to sequence RNAs {\em in situ}, simultaneously determining their identity and location. As a step along this path, we developed methods to cryosection individual blastoderm stage {\em Drosophila melanogaster} embryos along the anterior-posterior axis and sequence the mRNA isolated from each 25\micron{} slice. The spatial patterns of gene expression we infer closely match patterns previously determined by {\em in situ} hybridization and microscopy. We applied this method to generate a genome-wide timecourse of spatial gene expression from shortly after fertilization through gastrulation. We identify numerous genes with spatial patterns that have not yet been described in the several ongoing systematic {\em in situ} based projects. This simple experiment demonstrates the potential for combining careful anatomical dissection with high-throughput sequencing to obtain spatially resolved gene expression on a genome-wide scale. 


\section*{Introduction}

Analyzing gene expression in multicellular organisms has long involved a tradeoff between the spatial precision of imaging and the efficiency and comprehensiveness of genomic methods. RNA {\em in situ} hybridization (ISH) and antibody staining of fixed samples, or fluorescent imaging of live samples, provides high resolution spatial information for small numbers of genes \cite{Fowlkes:2008ca,Tomancak:2007dg,Lecuyer:2007bc}. But even with automated sample preparation, imaging, and analysis, {\em in situ} based methods are difficult to apply to an entire genomes worth of transcripts or proteins. High throughput genomic methods, such as DNA microarray hybridization or RNA sequencing, are fast and relatively inexpensive, but the amount of input material they require has generally limited their application to homogenized samples, often from multiple individuals. Methods involving the tagging, sorting, and analysis of RNA from cells in specific spatial domains have shown promise \cite{Steiner:2012gg}, but remain non-trivial to apply systematically, especially across genotypes and species.

Recent advances in DNA sequencing suggest an alternative approach. With increasingly sensitive sequencers and improved protocols for sample preparation, it is now possible to analyze small samples without amplification. Several years ago we developed methods to analyze the RNA from individual {\em Drosophila} embryos \cite{Lott:2011cc}. As we often recovered more RNA from each embryo than was required to obtain accurate measures of gene expression, we wondered whether we could obtain good data from pieces of individual embryos, and whether we could obtain reliable spatial expression information from such data. To test this possibility, we chose to focus on anterior-posterior (A-P) patterning in the early embryo {\em D. melanogaster} embryo, as the system is extremely well-characterized and the geometry of the early embryo also lends itself to biologically meaningful physical dissection by simple sectioning along the elongated A-P axis.

\section*{Results}

To test whether we could consistently recover and sequence RNA from sectioned  {\em D. melanogaster} embryos, we collected embryos from our laboratory stock of the line CantonS (CaS), aged them for approximately 2.5 hours so that the bulk of the embryos were in the cellular blastoderm stage, and fixed them in methanol. We examined the embryos under a light microscope and selected single embryos that were roughly halfway through cellularization (mitotic cell cycle 14; developmental stage 5). We embedded each embryo in a cryoprotecting gel, flash-froze it in liquid nitrogren, and took transverse sections along the anterior-posterior axis. For this initial trial we used 60\micron{} sections, meaning that we cut each approximately 350\micron{} embryo into six pieces. We placed each piece into a separate tube, isolated RNA using Trizol, and prepared sequencing libraries using the Illumina Tru-Seq kit .

In early trials we had difficulty routinely obtaining good quality RNA-seq libraries from every section. We surmised that we were losing material from some slices during library preparation as a result of the small amount (approximately 15ng) total RNA per slice. To overcome this limitation, after the initial RNA extraction we added RNA from a single embryo of a distantly related {\em Drosophila} species to each tube to serve as a carrier. As we only used distantly related and fully sequenced species as carriers, we could readily separate reads derived from the {\em D. melanogaster} slice and the carrier species computationally after sequencing. With the additional approximately 100ng of total RNA in each sample, library preparation became far more robust.

We sliced and sequenced three CaS embryos using an Illumina HiSeq 2000, obtaining approximately 40 million 50 bp paired-end reads for each slice+carrier sample. We aligned these reads to the {\em D. melanogaster} and carrier genomes using TopHat\cite{Langmead:2012jh,Kim:2013eo}, and identified between 1.7 and 31.4 percent of reads as having come unambiguously from {\em D. melanogaster} (see Table \ref{tab:seqstats}). We then used Cufflinks\cite{Roberts:2012kp} to infer expression levels for all annotated mRNAs. 

The data for each slice within an embryo were generally highly correlated (Figure S1), reflecting the large number of highly expressed genes with spatially uniform expression patterns. The data for equivalent slices of embryos 2 and 3 were also highly correlated, while the slices for embryo 1 were systematically less well matched to their counterparts in embryos 2 and 3 (Figure S2), suggesting that it may have been sampled at a slightly different developmental stage. 

\begin{figure}[!ht] 
\begin{center}
\includegraphics[width=6.93in]{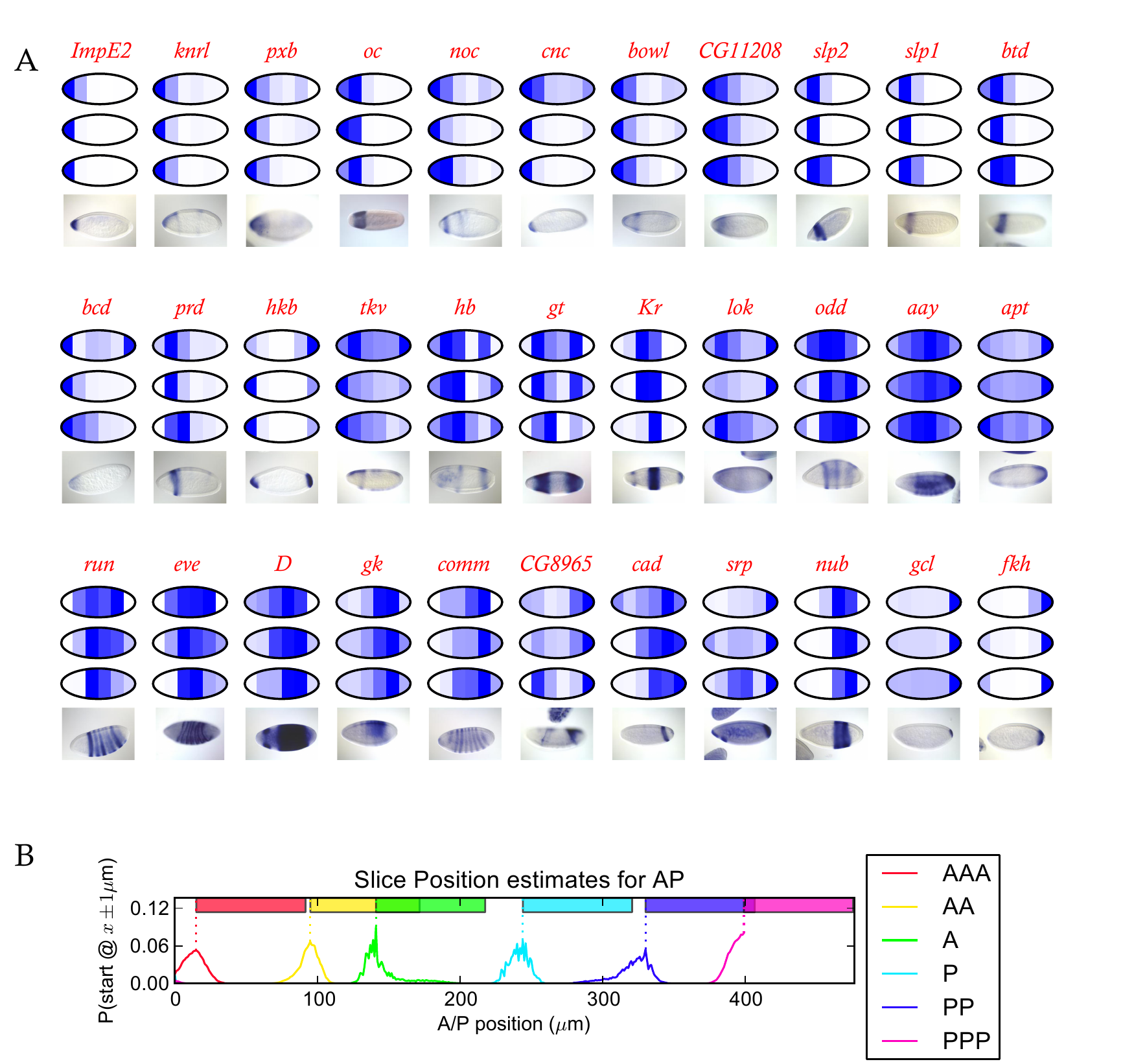}
\end{center}
\caption{ {\bf Expression in the slices closely matches previous expression data.} (A) 33 genes with previously known A-P patterns are shown with virtual {\em in situ} images for each of the three 60\micron{} sliced CaS embryos. The virtual {\em in situ} images are each scaled to the slice with the highest expression level for each embryo individually. (B) Expression data closely matches with previous quantitative data at the same stages.  We used a Bayesian procedure to estimate the location of each 60\micron{} slice with reference to an ISH-based atlas, with absolute expression levels set using whole embryo RNA-seq data.  The line graphs represent the distribution of position estimates for each slice, and the colored bars are one sixth the embryo width and placed at the position of greatest probability.}
\label{fig:posmatch}
\end{figure}

To examine how well our data recapitulated known spatial profiles, we manually examined a panel of genes with known anterior-posterior patterns of gene expression. Figure \ref{fig:posmatch}A shows RNA in-situ hybridization patterns from the Berkeley Drosophila Genome Project (BDGP) \cite{Tomancak:2007dg} alongside the expression data for that gene from our sliced embryos, demonstrating a close qualitative agreement between the visualized expression patterns and our sliced RNA-seq data. 

In order to more quantitatively compare our data to existing patterns, we constructed a reference set of spatial expression patterns along the A-P axis using three-dimensional ``virtual embryos'' from the Berkeley Drosophila Transcription Network Project, which contain expression patterns for 95 genes at single-nucleus resolution \cite{Fowlkes:2008ca}. We  transformed the relative expression levels from these images into absolute values (FPKM) using genome-wide expression data from intact single embryos \cite{Lott:2011cc}. We compared the observed expression for these 95 genes from an average of each of our slices to all possible  60\micron{} slices of these virtual embryos (Figure \ref{fig:posmatch}B). High scores for most slices fell into narrow windows, with the best matches for each slice falling sequentially along the embryo with a spacing of about 60\micron{}, the same thickness as the slices.

We next used the program Cuffdiff \cite{Trapnell:2012gg} to identify 85 genes with statistically significant differences in expression between slices (this is a very conservative estimate). We compared these genes to those examined by the BDGP, the most comprehensive annotation of spatial localization in {\em D. melanogaster} development that we are aware of \cite{Tomancak:2007dg}. Of our differentially expressed genes,  21 had no imaging data available, and 33 were annotated as present in a subset of the embryo (the annotation term meant to capture patterned genes); the remaining 31 genes showed either clear patterns that were not annotated with the most general keyword, or no clear staining (Figure S3).  There were 194 genes tagged by the BDGP as patterned that were not picked up as having statistically significant patterns in our data. However, most of these had primarily dorsal-ventral patterns, faint patterns, later staging in the images used for annotation, or had good qualitative agreement with our data but fell above the cutoff for statistical significance (Figure S4).

\begin{figure}[!ht]
\begin{center}
\includegraphics[width=6.93in]{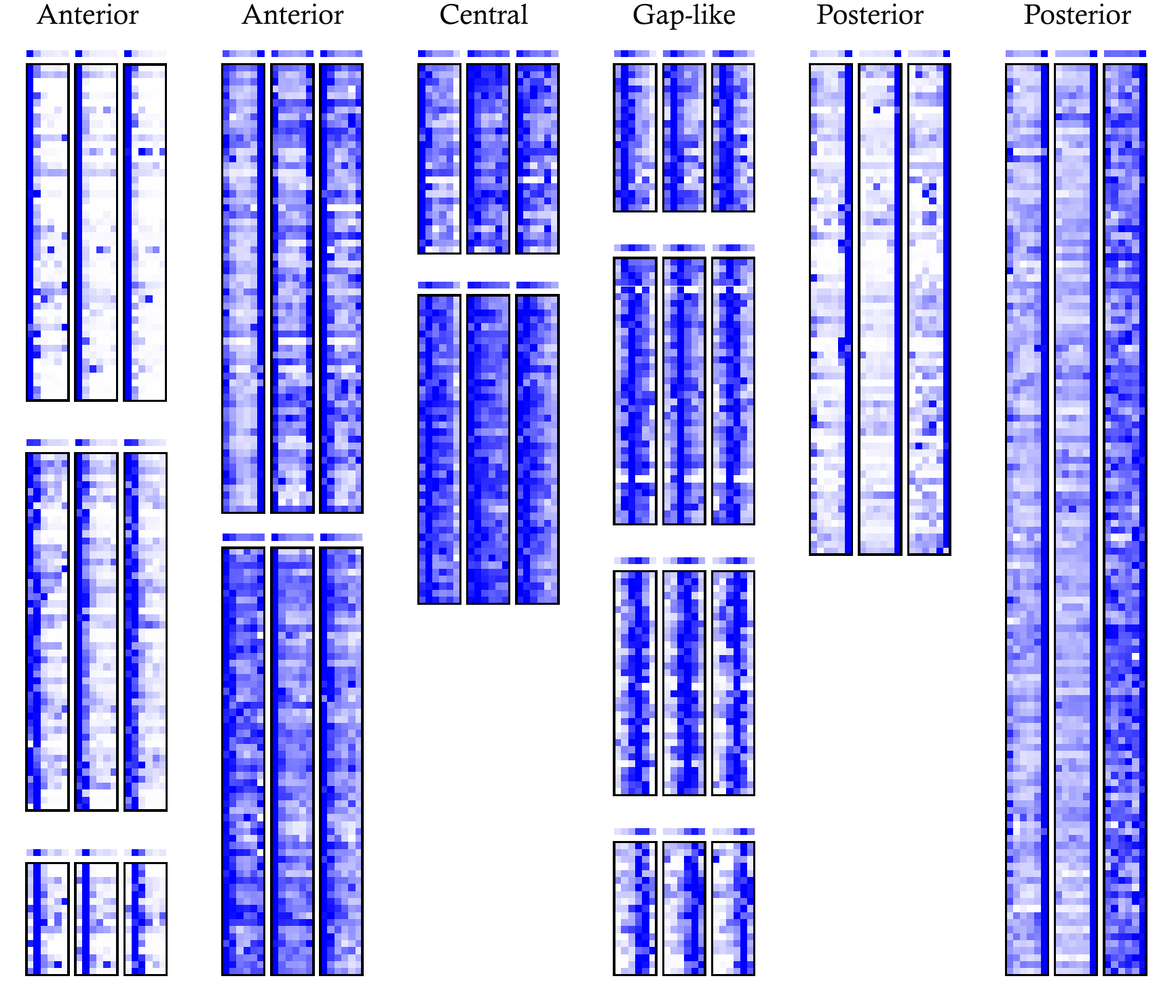}
\end{center}
\caption{
{\bf Heat maps of gene expression clusters}. Of the $k=20$ clusters, 13 with non-uniform patterns are shown. The expression levels for each gene was normalized for clustering and display so that the maximum expression of each gene in each embryo is dark blue. The plot above each cluster is the mean normalized expression level in that cluster.} 
\label{fig:clusters} 
\end{figure}

As a more sensitive approach to finding patterned genes, we applied $k$-means clustering to our data.  We first filtered on expression level (at least one slice in one embryo with FPKM > 10) and agreement between replicates (average Pearson correlation between embryos of > 0.5), then clustered based on normalized expression ($k=20$, centroid linkage) \cite{deHoon:2004gf}. We identified several broad classes of expression, including localization to each of the poles, and several classes of expression that correspond to five different gap gene-like bands along the AP axis\ref{fig:clusters} and Figure S5. Of the 745 genes, only 349 had images in the BDGP set\cite{Tomancak:2007dg}. Where present at similar stages, this data agrees with the RNA-seq patterns, although staining is often undetectable and well-matched stages are often missing from the databases (Figure S6).  

To extend our dataset, we collected individual embryos from seven different time points based on morphology---stage 2, stage 4, and 5 time points within stage 5---and sliced them into 25\micron{} sections, yielding between 10 and 15 contiguous, usable slices per embryo.  For these embryos we used total RNA from the yeasts {\em Saccharomyces cerevisiae} and {\em Torulaspora delbruckii} as carrier, which are so far diverged as to have fewer than 0.003\% of reads ambiguously mapping. 

\begin{figure}[!ht]
\begin{center}
\includegraphics[width=6.93in]{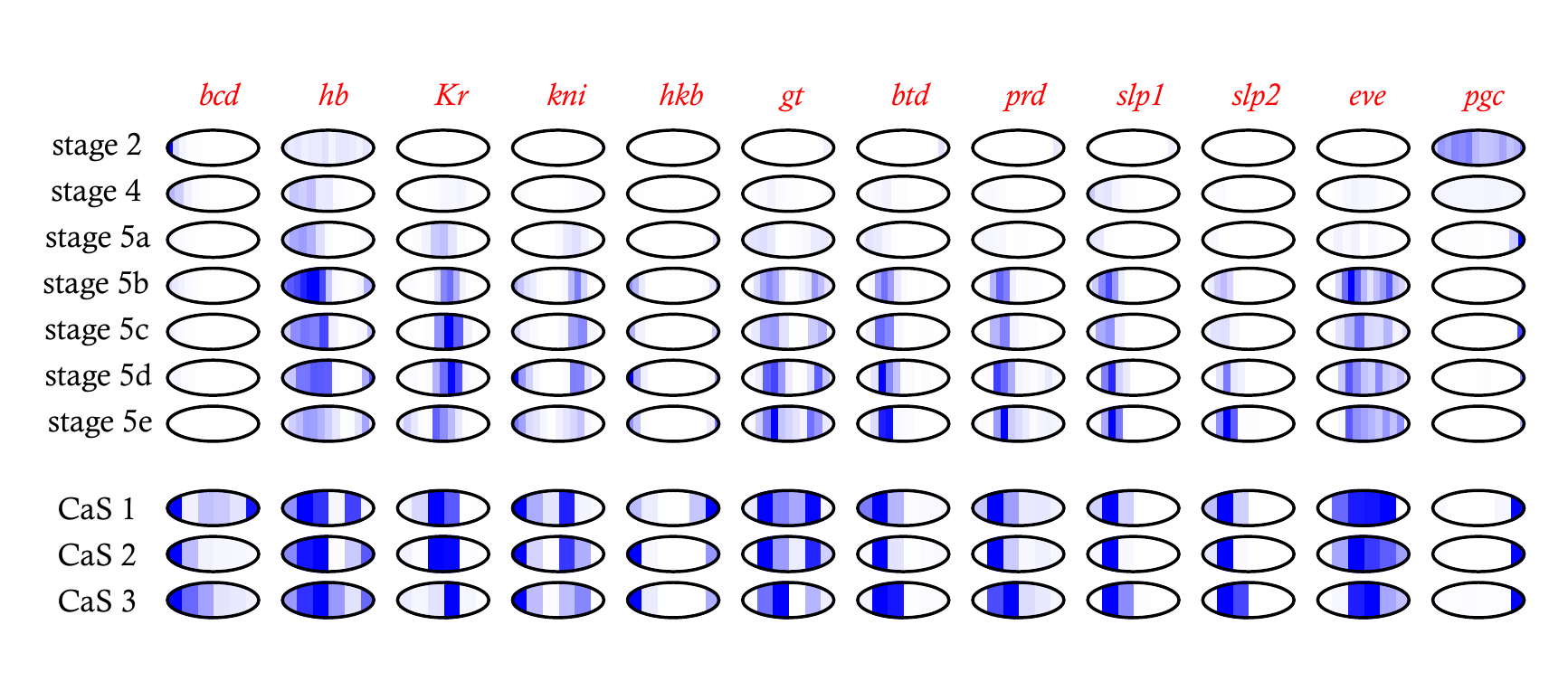}
\end{center}
\caption{
{\bf Expression of key patterning genes across early development.} Expression levels in the 25um timeseries are normalized to the highest expression level at any time pioint. For slices with poor quality data (timepoint 4, slice 10; timepoint 6, slice 6; timepoint 7, slice 7; and timepoint 7, slice 8) data imputed from neighboring slices is shown. Expression levels for the 60um slice samples are normalized to the highest level in each embryo. }
\label{fig:fineslice} 
\end{figure}

These finer slices are better able to distinguish broad gap-gene domains, with several slices of relatively low expression between the multiple domains of {\em hb}, {\em kni}, and {\em gt}, whereas the coarser slices only have one, or at best two slices. Excitingly, we can also distinguish the repression between stripes of pair-rule genes like {\em eve} as well (Figure \ref{fig:fineslice}). Given the non-orthogonal orientation of the anterior-most and posterior-most {\em eve} stripes relative to the AP axis, we do not expect to see all 7 pair-rule stripes, but at least three can be unambiguously observed. 

Putting the 60um and 25um slice datasets together, we find a large number of genes with reproducible patterns in the 60um slices whose formation over time can be clearly seen in the timed 25um slices, including many without no previously described early patterns (Figure S7).

\section*{Discussion}

The experiments reported demonstrate that slicing and sequencing animal embryos is a practical and effective method to systematically characterize spatial patterns of expression. While we are by no means the first to dissect samples and characterize their RNAs---Ding and Lipshitz pioneered this kind of analysis twenty years ago \cite{DaliDing:2004wu}---to our knowledge we are the first to successfully apply such a technique to report genome-wide spatial patterns in a single developing animal embryo. 

Given the degree to which the {\em D. melanogaster} embryo has been studied, and the presence of at least two large {\em in situ} based studies whose goals were to systematically identify and characterize genes with patterned expression in the embryo, we were surprised by the large number of genes we find as clearly patterned that had not been previously described as such. We note in particular a large number of genes with expression restricted to the poles, most with no known role in either anterior patterning or pole cell formation or activity. This emphasizes the potential for sequencing-based methods to replace {\em in situ} based studies in the systematic analysis of patterned gene expression, as they are not only simpler, cheaper, and easier to apply to different species and genetic backgrounds, but appear to be more sensitive. 

The data we present here are far from perfect - the relatively small number of reads per slice means that the slice by slice data are somewhat noisy. However the consistency between replicates and the agreement between the 25um and 60um data demonstrate that the experiment clearly worked, and additional sequencing depth and better methdos for working with small samples should greatly reduce the noise as we move forward. 

Obviously, to truly replace {\em in situ} based methods, sequencing based methods will need to achieve greater resolution than presented here. One can envision three basic approaches to achieving the ultimate goal of determining the location of every RNA in a spatially complex tissue. Sequencing RNAs in place in intact tissues would obviously be the ideal method, and we are aware of several groups working towards this goal. In the interim, however, methods to isolate and characterize smaller and smaller subsets of cells are our only alternative. One possibility is to combine spatially restricted reporter gene expression and cell sorting to purify and characterize the RNA composition of differentiated tissue---c.f. \cite{Steiner:2012gg}. While elegant, this approach cannot be rapidly applied to different genetic backgrounds, requires separate tags for every region/tissue to be analyzed, and will likely not work on single individuals.

Sectioning based methods offer several advantages, principally that they can be applied to almost any sample from any genetic background or species, and allow for the biological precision of investigating single individuals. The 60\micron{} and 25\micron{} slices we used here represent reasonable tradeoffs between sequencing depth and spatial resolution given the current limits of sample preparation and sequencing methods, but with methods having been described to sequence the RNAs from ``single'' cells, it should be possible to obtain far better linear spatial resolution in the near future.

Finally, as sequencing costs continue to plummet, it should be possible to sequence greater numbers of increasingly small samples.  According to our estimates, a single embryo contains enough RNA to sequence over 700 samples to a depth of 20 million reads. While this number of samples would necessitate more advanced sectioning and sample preparation techniques, the ultimate goal of knowing the localization of every single transcript is rapidly becoming feasible.

\section*{Materials and Methods}

\subsection*{Fly Line, Imaging, and Slicing}

We raised flies on standard media at $25^{\circ}$ in uncrowded conditions, and collected eggs from many 3 to 10-day old females from our {\em Canton-S} lab stocks.  We washed and dechorionated the embryos, then fixed them according to a standard methanol cracking protocol. Briefly, we initially placed embryos in 20ml glass vials containing 10ml of heptane and 10ml of PEM (100mM PIPES, 2mM EGTA, 1mM MgSO4) and mixed gently. We then removed the aqueous phase, added 10ml of methanol, shook vigorously for 15-30 seconds, and collected the devitellinized embryos, which we washed several times in methanol to remove residual heptane. We then placed the fixed embryos on a slide in halocarbon oil, and imaged on a Nikon 80i with DS-5M camera.  After selecting embryos with the appropriate stage according to depth of membrane invagination and other morphological features, we washed embryos with methanol saturated with bromophenol blue dye (Fisher, Fair Lawn NJ), aligned them in standard cryotomy cups (Polysciences Inc, Warrington, PA), covered them with OCT tissue freezing medium (Triangle Biomedical, Durham, NC), and flash froze them in liquid nitrogen.

We sliced frozen embryos on a Microm HM 550 (Thermo Scientific, Kalamozoo, MI) at a thickness of 60\micron{} or 25\micron{}.  We adjusted the horizontal position of the blade after every slice to eliminate the possibility of carry-over from previous slices, and used a new blade for every embryo.  We placed each slice in an individual RNase-free, non-stick tube (Life Technologies, Grand Island, NY).

\subsection*{RNA Extraction, Library Preparation, and Sequencing}

We performed RNA extraction in TRIzol (Life Technologies, Grand Island, NY) according to manufacturer instructions, except with a higher concentration of glycogen as carrier (20 ng) and a higher relative volume of TRIzol to the expected material (1mL, as in \cite{Lott:2011cc}).  For the 60um slices, we pooled total RNA from each slice with total RNA from single {\em D. persimilis}, {\em D. willistoni}, or {\em D. mojavensis} embryos, then made libraries according to a modified TruSeq mRNA protocol from Illumina.  We prepared all reactions with half-volume sizes to increase relative sample concentration, and after AmpureXP cleanup steps, we took care to pipette off all of the resuspended sample, leaving less than 0.5 $\mu L$, rather than the 1-3 $\mu L$ in the protocol.  Furthermore, we only performed 13 cycles of PCR amplification rather than the 15 in the protocol, to minimize PCR duplication bias.

Libraries were quantified using the Kapa Library Quantification kit for the Illumina Genome Analyzer platform (Kapa Biosystems) on a Roche LC480 RT-PCR machine according to the manufacturer's instructions, then pooled to equalize index concentration.  Pooled libraries were then submitted to the Vincent Coates Genome Sequencing Laboratory for 50bp paired-end sequencing according to standard protocols for the Illumina HiSeq 2000.  Bases were called using HiSeq Control Software v1.8 and Real Time Analysis v2.8.

\subsection*{Mapping and Quantification}

Reads were mapped using TopHat v2.0.6 to a combination of the FlyBase reference genomes (version FB2012\_05) for {\em D. melanogaster} and the appropriate carrier species genomes with a maximum of 6 read mismatches \cite{McQuilton:2011iq,Trapnell:2009dp}.  Reads were then assigned to either the {\em D. melanogaster} or carrier genomes if there were at least 4 positions per read to prefer one species over the other.  We used only the reads that mapped to {\em D. melanogaster} to generate transcript abundances in Cufflinks.

\subsection*{Data and Software}

We have deposited all reads in the NCBI GEO under the accession number GSE43506 which is available immediately.  The processed data, including a search feature of the 25\micron{} dataset, are available at the journal website and at eisenlab.org/sliceseq.  All custom analysis software is available github.com/petercombs/Eisenlab-Code, and is primarily written in Python \cite{VanRossum:2003tg,Cock:2009hj,Hunter:2007,Jones:2001uv,Perez:2007wf}.  Commit b0b115a was used to perform all analysis in this paper.


\section*{Acknowledgments}

We thank all who have contributed feedback through the open review of the manuscript on MBE's blog and the arXiv.

\bibliography{papers}

\clearpage

\section*{Figure Legends}


\clearpage
\clearpage

\section*{Supplemental Figure Legends}
Prepublication Supplemental Figures available by request.

{\bf Figure S1. Correlation of slices within embryos}. Log-log plots of FPKM values between slices within each of the three 60\micron{} sliced embryos.

{\bf Figure S2. Correlation of slices between embryos}. Log-log plots of FPKM values of corresponding slices between each of the three 60\micron{} sliced embryos.

{\bf Figure S3. Genes called as patterned by Cuffdiff lacking subset tag in BDGP database}. Images are from BDGP; graphs are average of three CaS embryos. Many of these are known patterned genes, highlighting the incompleteness of available annotations. 

{\bf Figure S4. Genes with subset tag in BDGP not called as patterned by Cuffdiff}.

{\bf Figure S5. Figure 2 with gene names}.

{\bf Figure S6. Images from BDGP for genes in clusters shows in Figure 2}. 

{\bf Figure S7. Data from 25\micron{} timecourse and 60\micron{} embryos for a large number of genes with manually curated patterns}.

\clearpage

\section*{Tables}

\begin{table}[!ht]
\caption{
\bf{Sequencing statistics for sliced single-stage wild-type mRNA-Seq samples}}
\begin{tabular}{|c|c|p{.5in}|p{0.5in}|c|p{1.15in}|p{1.0in}|}
	\hline
	Replicate & Slice & Carrier Species & Barcode Index & Total Reads
	& Uniquely mapped {\em D. mel} reads (\%) & Ambiguous Reads (\%) \\
	\hline
1 & 1 & {\em D. per} & 1  & 69,339,972 & 2,284,228 (3.2\%)   & 1,634,055 (2.3\%)\\
1 & 2 & {\em D. per} & 2  & 73,632,862 & 3,706,630 (5.0\%)   & 1,603,444 (2.1\%)\\
1 & 3 & {\em D. per} & 3  & 82,076,328 & 6,002,034 (7.3\%)   & 1,774,485 (2.1\%)\\
1 & 4 & {\em D. per} & 4  & 73,437,708 & 6,401,565 (8.7\%)   & 1,592,665 (2.1\%)\\
1 & 5 & {\em D. per} & 5  & 75,922,812 & 4,951,178 (6.5\%)   & 1,559,097 (2.0\%)\\
1 & 6 & {\em D. per} & 6  & 78,623,784 & 1,355,079 (1.7\%)   & 1,574,067 (2.0\%)\\\hline
2 & 1 & {\em D. wil} & 7  & 59,813,036 & 4,066,295 (6.7\%)   &  878,476 (1.4\%)\\
2 & 2 & {\em D. wil} & 8  & 90,961,338 & 15,212,716 (16.7\%) & 1,301,095 (1.4\%)\\
2 & 3 & {\em D. wil} & 9  & 73,201,902 & 14,855,374 (20.2\%) &  911,768 (1.2\%)\\
2 & 4 & {\em D. wil} & 10 & 75,754,772 & 23,858,301 (31.4\%) & 1,136,031 (1.4\%)\\
2 & 5 & {\em D. wil} & 11 & 84,497,566 & 10,026,713 (11.8\%) & 1,080,910 (1.2\%)\\
2 & 6 & {\em D. wil} & 12 & 66,316,952 & 13,122,508 (19.7\%) &  898,776 (1.3\%)\\\hline
3 & 1 & {\em D. moj} & 13 & 75,847,986 & 12,496,248 (16.4\%) & 3,615,452 (4.7\%)\\
3 & 2 & {\em D. moj} & 14 & 72,497,660 & 4,005,714 (5.5\%)   &  803,381 (1.1\%)\\
3 & 3 & {\em D. moj} & 15 & 77,532,368 & 11,138,154 (14.3\%) &  772,446 (0.9\%)\\
3 & 4 & {\em D. moj} & 16 & 83,400,882 & 8,227,562 (9.8\%)   &  861,839 (1.0\%)\\
3 & 5 & {\em D. moj} & 18 & 83,608,454 & 2,630,069 (3.1\%)   &  795,169 (0.9\%)\\
3 & 6 & {\em D. moj} & 19 & 85,823,784 & 2,239,493 (2.6\%)   &  829,382 (0.9\%)\\
\hline
\end{tabular}
\begin{flushleft}
	Counts are for read ends.
	Discordant read ends
	are always classed as ambiguous,
	but failure of one end to map
	does not disqualify the other.
\end{flushleft}
\label{tab:seqstats}
 \end{table}

\clearpage

\end{document}